# Assessing "Information Quality" in IoT Forensics: Theoretical Framework and Model Implementation*

Federico Costantini
*Researcher at the Department of Law*
*University of Udine (IT)*
`federico.costantini@uniud.it`

Fausto Galvan
*Digital Forensics Expert and Law Enforcement agent*
*Public Prosecutor office at the Court of Udine (IT)*
`galvanfausto14@gmail.com`

Marco Alvise De Stefani
*CEO of Synaptic Srls and Digital Forensics Expert*
`destefani@synaptic.it`

Sebastiano Battiato
*Full Professor of Computer Science*
*Department of Mathematics and Computer Science*
*University of Catania (IT)*
`battiato@dmi.unict.it`

**Abstract**

IoT technologies pose serious challenges to digital Forensics. The acquisition of digital evidence is hindered by the number and extreme variety of IoT items, often lacking physical interfaces, connected in unprotected networks, feeding data to uncontrolled cloud services. In this paper we address "Information Quality" in IoT Forensics, taking into account different levels of complexity and included human factors. After drawing a theoretical framework on data quality and information quality, we focus on forensic analysis challenges in IoT environments, providing a use case of evidence collection for investigative purposes. At the end, we propose a formal framework for assessing information quality of IoT devices for Forensics analysis.

---

*Although this contribution has to be considered as an outcome of an ongoing joint research, the single paragraphs should be mainly attributed as follows: 1-3 to Federico Costantini, 4 to Marco Alvise De Stefani , 5 and 6 to Fausto Galvan, 7 to Sebastiano Battiato.

**Keywords:** *Information Quality Assessment, Digital Forensics, IoT Forensics, Digital Investigations.*

# 1  Introduction: challenges in IoT Forensics

In the last twenty years, we have been witnessing the advancement of a forensic science known as Digital Forensics, whose aim is to develop a rigorous methodology for retrieval, collection, and analysis of digital evidence[1]. One of the greatest challenges in this discipline is to keep up with the speed of technical innovation that, as we know, in the digital field is higher than in any other sector. In recent years, indeed, the analysis of such evidence has required demanding effort, especially due to the pervasive use of two technologies: cloud computing and artificial intelligence. On the one hand, the virtualization of resources hinders the validation of the source, the accuracy of the analysis, and the integrity of the results, since «*evidence can reside everywhere in the world in a virtualization environment*» [68]. On the other hand, the background of a decision taken by artificial intelligence systems lacks transparency, in that these systems' behaviour is as unpredictable as "black box" outcomes.

Currently, forensic analysis of such systems has been addressed in an effort to improve pre-existing methodologies. "Cloud Forensics" can be defined as «*the application of Computer Forensics principles and procedures in a cloud computing environment*» [59] whereas "explainable artificial intelligence" (XAI) aims to develop a suite of techniques that, bringing more transparency into the reasoning process, allows the validation of the reconstruction of external events [1]. Despite the promising results currently achieved in this way, such "disruptive technologies" require not only new technologies but also new approaches which have not been completely deployed yet.

We can observe the same evolution in the Internet of Things (IoT). The term was originally coined in 1999 with specific reference to RFID (Radio-Frequency

---

[1]Digital Forensics has been defined as "*the use of scientifically derived and proven methods toward the preservation, collection, validation, identification, analysis, interpretation, documentation, and presentation of digital evidence derived from digital sources for the purpose of facilitation or furthering the reconstruction of events found to be criminal, or helping to anticipate unauthorized actions shown to be disruptive to planned operations.*" [58].

Identification) technologies [4] and soon overcame more general expressions such as "ubiquitous computing" [73], "pervasive computing" [34] and "ambient intelligence". Now it is commonly used to designate «*a global infrastructure for the information society, enabling advanced services by interconnecting (physical and virtual) things based on existing and evolving interoperable information and communication technologies*», according to a definition given by ITU in 2012 [45, 30]. A previously only imagined scenario now characterizes everyday life. Today millions of devices – of many different types, models, and versions – are connected in an extensive infrastructure, exchanging enormous quantities of data, and it can be foreseen that this trend will increase dramatically in the future. According to a report by Ericsson, there could be 3.5 billion IoT connections in 2023, 2.2 billion of them in North-East Asia alone [17].

Forensic analysis of digital evidence in the IoT environment poses several challenges [35, 36, 80, 54]. Indeed, data are spread across an undetermined (e.g. in their type, number, and location) set of connected devices [79]; machines present different kinds of security vulnerabilities, so in being extensively exposed to hijackers, communications can be unprotected or even unencrypted, allowing third-party manipulation; and storage units often do not grant secure access to archives. Furthermore, due to the high interdependence among devices, any anomaly can spread rapidly in an IoT ecosystem and flood outwards; hence serious crimes, with great damage of exponential scale, could be committed without leaving any trace. As a matter of fact, methods tested as valid for isolating devices in "chain of custody"[2] situations, as in "classical" digital Forensics, are not effective in the IoT environment, due to the constant connections and deep interaction among devices.

IoT Forensics analysis requires both cutting-edge technological solutions and new methodological approaches in order to grant integrity, authentication, and non-repudiation of digital evidence[3] extracted from the acquired data. Three specific reasons can support such claims: objective, subjective, and transactive.

First, objective: the greater the quantity and diversity of evidentiary sources, the less trust can be put on their quality. With a huge amount and variety of sources to analyse, there is a great impact on the extraction of data. If many different kinds of devices are interconnected, it is likely that they must be approached with different standards and, conversely, it is unlikely that the results can be efficiently aggregated.

Second, subjective: the more data that are available, the harder it is to manage them, even if deploying cutting-edge Forensics tools. Indeed, the quantity of evidence has a considerable influence on the computational effort required for information retrieval. Selection of relevant information and the processing of results can be very demanding, also considering that, especially in ongoing trials, the usefulness of the outcome strictly depends on its timeliness, and the analysis of an IoT ecosystem requires a consistent effort by Forensics consultants.

---

[2]The set of procedures aiming to guarantee the integrity of evidence is called "chain of custody". It encompasses different phases and requires operations to be logged in order to ensure transparency and trust. See the "Convention on Cybercrime" signed in Budapest in 2001 (n. 185 CoE).

[3]A new and interesting perspective is brought by decentralized ledger technologies, as recently pointed out in an interesting paper by Brotsis [14].

Third, transactive: the more advanced the technologies, the higher the expertise required for their analysis, and the lower the average understanding capacity by non-experts. Digital Forensics, indeed, as with any other forensic discipline, requires not only a solid background, rigorous methodology, professional experience, competences, skills, and up-to-date technological tools, but also the ability to support findings with convincing arguments. Evidence, even of a digital nature, needs to be not only acquired but also discussed with interlocutors who are not experts in the specific fields: judges, police officers, lawyers, and other consultants. If not plainly explained, the most grounded evidence might have no impact even if data were to be perfectly available, accessible, and genuine.

Digital Forensics analysis, and thus IoT Forensics, cannot be addressed without considering the three issues described above, which encompass the general problem of the influence of human factors in the transmission of knowledge. Of course, any forensic discipline is grounded in scientific methodology, and its assumptions are commonly justified by technological standards, established protocols, and widely accepted best practices. Moreover, the admission of evidence in court is ruled by adjective procedures, which confer strict obligations on all parties (judges, prosecutors, defendants, police officers, witnesses, consultants). Even so, it still is important to underline that the purpose of evidence is to be discussed and decided upon. In other words, both validity and efficacy of proof are noteworthy. The fact that evidence is digital, thus involving specific technologies and expertise, should not be allowed to undermine the issues concerning their impact in courts.

In this paper we explore IoT Forensics in terms of information quality (henceforth, IQ), offering a theoretical model which includes both data quality (henceforth, DQ), technological resources and human aspects. To that aim, we intend to proceed as follows: firstly, we provide an overview on DQ and IQ concepts, focusing on IoT issues; secondly, we focus on the concept of IQ we intend to explore in this contribution, including it in a wider theoretical framework; after that, we analyse the practical concerns, expressing the issues pointed out in a previous part; consequently, we propose a model for assessing IQ in the IoT environment and represent it in a formula. In conclusion, we express our final evaluations and draw paths for future research.

## 2 Data, Information, DQ and IQ: an Overview

"Data" and "information" are two deeply intertwined concepts whose core and mutual difference have been extensively debated among different approaches and disciplines. On one side, a pragmatic view can be epitomized by technological standards. In this sense, according to [39] "data" is defined in terms of "information", while "information" is considered a sort of "knowledge"[4]. Conversely, a remarkable example of a theoretical perspective can be found in "Philosophy of information", an approach developed in last the twenty years [25, 26], according to which "information" is described as a sort of "meaningful data" while "knowledge" is theorised as "accounted information" [28]. This latter proposal allows to shape a thorough vision which encompasses traditional ontology and contemporary epistemology and it is able to include also the social processes of transmission and sharing. In other terms, reality and knowledge are perceived and converted in "meaning" by agents – if humans, through interpretation, if machines, thanks to computation[5] – and then shared among each other regardless of their organic or artificial nature. In this view, we, as humans, are informational agents populating an ecosystem called "infosphere" as well as animals and computers [27].

In the legal field, the intricate relationship between reality and knowledge has always challenged the acquisition and admission of evidences. Indeed, in criminal proceedings every legal system entitles authorities with investigative powers for the collection of sources of evidence, which are submitted to judges according to specific procedures in order to bring decisions based upon them. Likewise, evidences are debated in civil trials where parties are equally enabled – at least in general – to produce facts or circumstances in support to their arguments or counter-arguments. Provided that an evidence – regardless the fact that it is incorporated in a digital support or not – can be qualified a sort of "information", the problem of IQ becomes crucial for the application of law. In other words, it is important to establish transparent standards for the acquisition and admission of evidences in order to create an environment suitable for celebrating fair trials. This is essential in the case of digital forensics – especially in IoT forensics, as anticipated in the introduction – since the evidence is electronic, thus not engraved in a physical object.

---

[4]Indeed, "data" is defined as: "reinterpretable representation of information in a formalized manner suitable for communication, interpretation and processing", while "information" is defined as: "information processing knowledge concerning objects, such as facts, events, things, processes, or ideas, including concepts, that within a certain context have a particular meaning" [8].

[5]In this paper we cannot delve the difference between interpretation and computation. Our intention is to draw an abstract model suitable to describe the process made by the human mind and by an artificial agent.

## 2.1 Data and DQ

It has been said that data are the bridge that links the physical realm and the cyber world [48]. The life cycle of data has evolved due to recent technological innovations, as information is now processed both by humans and machines. Since IoT technologies allow an extensive, continuous, and direct flow of information among devices, the problem of DQ is crucial, especially if the interaction is not filtered by human supervision. Provided that the IoT is intrinsically untrustworthy due to the reasons mentioned above, this can lead to unpredictable consequences of an exponential scale, but the IoT also can generate anomalies which are also unperceivable to users, especially humans, thus hampering the possibility of arranging countermeasures or remedies.

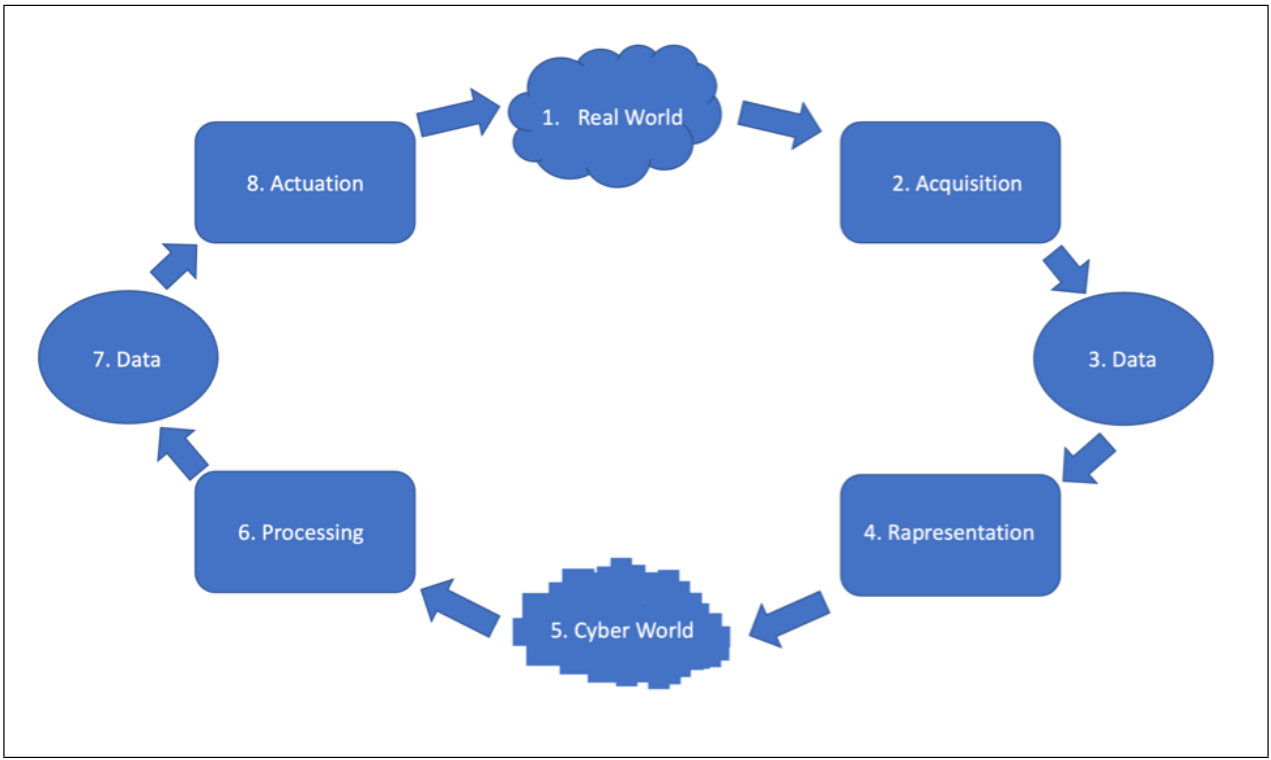

Figure 1: Representation of IoT scenario

As shown in Fig.1, the pipeline that allows one to collect a reliable set of information from a general IoT scenario starts with the acquisition of a set of data from the "real world". This could be done by a group of sensors, or by automated collecting procedures provided by the data owner, depending upon their data policy. This set of data is represented to the "cyber world" and then processed and re-introduced into the environment.

Although the concept of DQ has been crystalized by technical standards, being defined in ISO/IEC 25012:2008 [39] "*the degree to which the characteristics of data satisfy stated and implied needs when used under specified conditions*"[6], there is wide debate over the scope and reference of the "fitness for the (intended) use of data" [21, 22, 7, 75][7]. Specifically, scholars have discussed the classification of such feature, among which some 159 components have been counted [48]. Table 1[8]. shows a comparison of the different classifications.

In this paper, we make a methodological choice, preferring IQ over DQ in our approach, for three basic reasons: (1) the extreme complexity of IoT environments, as described above, in which data are just a part of a complex pipeline, along with devices and processes; (2) the possibility not only to include different kind of content – such metadata – but also to take into consideration the context of the devices, thus enabling a better representation of the complexity in the analysis; and (3) the possibility to include human interaction in the analysis of forensic evidence, since the meaning of data becomes as relevant as their source and the process performed to obtain it.

## 2.2 Information and IQ

Since "information" has a broader meaning than "data", as we argued above, the scope of IQ is wider than of DQ. As for the latter, in IQ scholars have proposed different criteria of classification – distribution, heterogeneity, and autonomy – which allow one to establish six different types of information systems (monolithic, distributed, data warehouses, cooperative, cloud, and peer to peer) [8]. Yet, one of the most interesting features of information is that it can be directly connected to the quality of the decisions that are based upon it. In this sense, an agent – either human or artificial – is influenced not only by shortage or by overload of information, but also by its quality. IQ, in short, is crucial for the outcome of the process, that is the utility of the decision in itself. Under this view, we can observe that some of the issues in IQ are the same as in DQ, for example in defining the dimensions under which quality can be addressed. In the field of Forensics analysis, two further profiles come into consideration if we consider the interaction between the devices examined and the Forensics expert. Indeed, the latter is entitled to bring the Forensics analysis to court and thus embodies the human decision maker for the filtering of data. Specifically, we have to consider (1) how information is collected by the decision maker and (2) how information affects the decision. Concerning the first aspect, it is noteworthy that IQ represents a kind of "meta-information", which should be taken into consideration by the agent in the decision. Here two aspects need to be measured: complacency[9] and consensus[10]. With regard to the second aspect, not only does the interpretation of information come into play (the accuracy and the depth of understanding of the agent) but also pre-existing factors – such as the agent's beliefs – or contextual aspects such as cognitive overload or time constraints[11].

---

[6]As stated by Batini and Scannapieco in [8], quality in general has been defined as the "totality of characteristics of a product that bear on its ability to satisfy stated or implied needs" ISO 9000:2015 [38]. See also the definition of "data quality" in ISO/TS 8000-1:2011 [44]

[7]Also called "fitness for (intended) use" [47]; "conformance to requirements" [20], or "user satisfaction" [71]

[8]The first four columns of the table represent different classifications proposed by four contributions mentioned in [8] which did not include [67], probably because it was published afterwards. Other models mentioned by Batini and Scannapieco are not tackled in this contribution being not relevant for IoT forensics [46, 12, 50, 56, 61]

| (Batini and Scannapieco 2016) | | | | |
|---|---|---|---|---|
| (Batini, Palmonari, and Viscusi 2012) | Theoretical approach (Wand and Wang 1996) | Empirical approach (Wang and Strong 1996) | | (Shamala et al. 2017) |
| Accuracy | Accuracy | Intrinsic | Accuracy | Accuracy |
| - | - | | Believability | Believability |
| - | - | | Objectivity | - |
| - | - | | Reputation | - |
| Completeness | Completeness | Contextual | Completeness | Completeness |
| - | - | | Relevancy | Relevancy |
| - | Timeliness | | Timeliness | Timeliness |
| - | - | | Value added | - |
| Redundancy | - | | - | - |
| - | - | | Appropriate amount of data | Amount of data |
| - | - | Representational | Interpretability | - |
| - | - | | Ease of understanding | - |
| - | - | | - | Objective |
| - | - | | Concise representation | Concise representation |
| Consistency | Consistency | | Representational consistency | Consistent representation |
| Usefulness | - | | - | - |
| Trust | - | | - | - |
| - | Reliability | | - | Reliability |
| Accessibility | - | Accessibility | - | Availability |
| - | - | | Accessibility | - |
| - | - | | Access security | - |

Table 1: Comparison of different classifications of features in DQ.

## 3  Theoretical background of IQ in IoT Forensics

According to "Philosophy of Information", as outlined above, "information" has three ontological statuses, which take into account the seminal studies of Weaver[12]: "information *as* reality", for example the electrical signal, which is transmitted regardless of the message contained; "information *about* reality", such as information about natural phenomena, which can be true or false (hence in philosophical terms can be said to be "alethic") [51, 23]; and "information *for* reality", which conveys instructions or algorithms to one or many recipients. Each of them has to be considered separately, since they refer to different aspects of reality.

Therefore, IQ can be studied under three different views.

- **Quality in "information *as* reality"** is the most common feature and emerges for example in the traditional problem of reducing noise, distortion, or losses in signal transmission. In this sense, it measures the affordability of the means implemented to transfer information;

- **Quality in "information *about* reality"** is concerned with the dissimilarity of information to the facts to which it refers. This concept is related to the meaning of information, or semantics [10, 6]. Thus, quality in this case measures the reliability of the information provided in representing the related events;

- **Quality in "information *for* reality"** has to be addressed when processes present inconsistencies, loopholes, or conflicts. This concept involves further processes of information, for example when it is shared with others. Hence, in this respect it measures the trustworthiness of the agent who receives information or, generally, of those involved in further processes.

---

[9] This feature measures "*the degree to which information on IQ is ignored*" [8].

[10] This feature measures "*the level of agreement within a group with respect to a preferred choice*" [8].

[11] These two aspects correspond respectively to the distinction between "*the purpose/s for which some information is originally produced (P-purpose) and the (potentially unlimited) purpose/s for which the same information may be consumed (C-purpose)*" [25]

[12] In the original exposition of the theory of communication, such concepts were expressed as different "levels", respectively as "technical", "semantic", and "influential" [72]. Instead, cybernetics defined "technological", "natural", and "cultural" information [11].

In Digital Forensics, the problem of IQ has received special consideration under different perspectives. Here we can deploy the same tripartite analytical model to draw a comprehensive framework.

- The first kind of quality is relevant in order to preserve the integrity of the collected information, since experts developed the concept of chain of custody and achieved a broad consensus on best practices which have been codified in technological standards[40, 41, 42, 43];

- The second type of quality is concerned about the trustworthiness of the representation of events, which has to be verified with other sources of evidence[29];

- The third sort of quality is involved in the discussion of evidence among parties (inquiring authorities, defendants, judges, Forensics experts). As we know, trials have to proceed according to precise rules which establish specific requirements for admissibility and the burden of proof. Here also, external variables can make a difference, such as personal competences of the agents involved, "soft skills" (argumentation abilities, trial strategies), cost of analytical tools, and available time.

From a theoretical perspective, since devices in the IoT are constantly sharing information with each other, IQ is a highly complex issue. Indeed, it has to be considered not only as a property of a single device but as a feature of the whole ecosystem in which every item is immersed. Specific concerns emerge in each considered aspect:

- **Quality of "information *as* reality"**. It is difficult to isolate a single device or crystalize a specific piece of information. The boundaries of relevance are blurred. This aspect is, not without reason, widely discussed by experts[74, 18];

- **Quality of "information *about* reality"**. It is problematic to detect a specific source, to trace the chain of interactions, or to measure the influence of a single item in shaping the representation of an event. It is commonly true that correlation is not causation; however, the IoT is, above all, a matter of correlation. Here in the IoT is where quality of information really faces uncertainty[52];

- **Quality of "information *for* reality"**. This is the most difficult aspect of IoT Forensics. Under this perspective, the human factor plays a part along with technological variables, as shown in digital Forensics.

| Ontological statuses of information | Quality of information; level of analysis | Quality of information in digital Forensics | Quality of information in IoT Forensics |
|---|---|---|---|
| Information *as* reality | Traditional theory of communication | Chain of custody | Relevance |
| Information *about* reality | Consistency with other represented facts | External validation with other sources of evidence | Uncertainty |
| Information *for* reality | Logical coherence | Adjective rules (admission and burden of proof) | Accountability |

Table 2: IQ tripartite analysis and IoT issues.

The classification outlined above is represented in Table 2.

Accountability is noteworthy in social processes, hence its legal relevance. This concept finds its roots in management theory, where one can find this general definition: *"to be accountable for one's activities is to explicate the reasons for them and to supply the normative grounds whereby they may be justified"* [33][13]. Accountability is crucial in fiduciary positions held on behalf of third parties, which are not directly involved in decisions that an agent has to make. The third party has the power to set a certain policy under which decisions have to be made by the agent, who is required not only to act according to said policy but to explain the reasons for her/his choices [49].

## 4  Practical issues in IoT Forensics

Accountability is fundamental in Digital Forensics, especially when solutions cannot be granted with certainty. When opinions are debatable, it is very important to be transparent about the methods adopted and to conveniently share the results obtained.

For example, during the bitstream copy of a hard disk it is very important to be transparent about the handling of the hard disk, the write-blocker and imaging software used, the log of the imaging process, the hash of the image, and so on; with this transparency, we know how the process is performed and thus can assess the IQ of the results obtained.

In IoT Forensics the observer faces three main problems related to accountability. She or he or needs to:

---

[13]In corporate management studies, accountability is crucial in decisions by corporate boards [64]

1. Acquire only data that are related to the case and that can be of interest. Otherwise, in IoT it is very easy to flood a case with billions of data points that after analysis come out to be useless. Selection has to be justified;

2. Assess the degree of uncertainty of the information, which must be a genuine representation of the reality. Such evaluation has to be clarified in its tenets;

3. Choose the tools of acquisition. Such a choice has to be explained, moreover, if for technical reasons the acquisition cannot be repeated in the future under the same conditions.

The main feature of an IoT network is the possibility, while performing a given tasks, and when necessary, to take advantage of information exchange with other neighbouring elements, regardless of the fact that they could belong to a different IoT ecosystem, possibly set up for a different purpose. Over the years it has become clear that, while it was originally conceived as the beginning of new opportunities for increasing the efficiency of services – thus to benefit private users and industries – this technology now generates a "virtual environment" containing a huge amount of information which, if needed, can be used as a source of evidence in a forensic scenario. In the following part, we observe that the main task of IoT Forensics, namely the collection of evidence, is challenging for different reasons.

First, the extreme variety of IoT items, with proprietary or undocumented protocols, often without physical interfaces, hinders the direct extraction of evidence from devices.

Second, IoT devices ceaselessly send information to "their own" cloud service providers. Thus, obtaining evidence from a cloud service involves other challenges, both technical (cloud Forensics) and legal (interacting with foreign companies in different legal frameworks).

Third, evidence could be manipulated by cloud service providers. And raw data fed by IoT devices are analysed, parsed, and stored in databases and servers, adding yet more layers of interpretations and classifications. Those processes can aggregate data but also deteriorate them, thus weakening the quality of the acquired information.

To explain such issues, we provide an example. Let's assume that we need to acquire the complete geolocalization history of a specific account, from a cloud service provider (i.e. Google Timeline History[14]).

Let's suppose also that we have the account's credentials (i.e. username and password): we can use forensic software (i.e. Oxygen Forensic Cloud Extractor[15]) to acquire the geolocalization history[16]. After the acquisition, we can explore and analyse the collected data: we will find a very clear and detailed set of information, spanning years and neatly organized. The typical forensic software user interface for this type of data consists of a world map we can freely zoom and search, a time filter, and coloured pins and lines that represent geolocalization information and probable movements.

---

[14] https://www.google.com/maps/timeline.
[15] https://www.oxygen-forensic.com/en/oxygen-forensic-cloud-investigator-ofci

This variety of sources can lead us to assume that the quality of data is very high, but after an additional analysis we may find that we don't know where the information comes from. We know that the service provider could acquire geolocalization information from a great range of sources related to that specific account: a smartphone's GPS; EXIF data of images; a cellular network radio tower; WiFi known geolocalization; near-field communication; fitness apps and arm bands; Uber rides; IP geolocalization; smartwatches; navigation apps; reviews of shops or restaurants; public transportation tickets; and many other IoT sources. The timeline can even be manipulated by the user through the cloud service web interface (i.e. in Google Timeline we can easily add a new geolocalization or change or delete an old record).

If we focus on a specific day, we only have the alleged geolocalization history: some points (consisting of longitude, latitude, and time stamp) and lines that connect those points to speculate movements, but we are not able to discover where each point comes from. Even if in some metatag there is information about the source of the info, we can access only the parsed data; we don't know if the source is reliable or if the data has been manipulated, compromised, or not properly interpreted. The original data may not even be accessible (i.e. if the geolocalization comes from a fitband's GPS, probably we can't physically connect to the device; we can acquire the paired smartphone and analyse the fitband's app's databases, but even that isn't the original data).

In this specific example (Google) we can also acquire the same data through the Google Takeout procedure in two different formats: JSON and KML.

We can try to understand what happened in reality only by comparing these three sources (Timeline and JSON/KML Takeout) and interpret the results with the experience and the reverse engineering in similar cases[17].

---

[16]We presume that the forensic software is compliant with all digital Forensics procedures and standards

[17]In other words, the same sporadic geolocalization (down to the cm) on different days, with a large uncertainty range, probably means that the smartphone was connected to a cellular radio tower

# 5 Proposal for a theoretical model of Information Quality Assessment in IoT Forensics

The evaluation of IQ in a set of heterogeneous data is not a trivial task, as the previous sections have shown. The various steps of the assessment always hang on different factors, and many of these cannot always be precisely quantified but only be estimated[18]. In addition to these basic difficulties, in common for every environment, as said before the IQ needed in a Forensics scenario must undergo a further filter, represented by a "check for admissibility", which allows the acquired data to be part of a trial procedure.

## 5.1 From the theoretical framework to a mathematical approach

A possible approach for such an issue could be based on the separate verification of IQ in two different fields, as shown in Figure 2. The information must be proved to be at the same time admissible as evidence in a trial – thus properly acquired and stored according to the latest and best practices in the field [40] – and reliable as a collection of data (with a low amount of noise, readable, understandable, saved in a known format, etc.); thus it will serve as input for the Forensics analysis. This method, although interesting, is affected by two main risks: wasting time with one of the two controls when the outcome of the other is negative, and rejecting key evidence which, for some reason, is not proven to be reliable as high quality IoT data (e.g. a noisy image which, considered together with other evidence, could have an important meaning).

This last observation leads to the development of another approach, where the two aspects of the extracted information (since they are, at the same time, "rough" data and "evidence") are spread all over the items of a sum of proper features.

Following the above considerations, we implemented the theoretical framework drawn in the previous section concerning the classification of the requirements of IQ in an IoT environment. For sake of clarity, we consider only the categories of IQ features shown in [69].

In our fictional investigative scenario, all the collected digital evidences come from a set of n IoT devices. Our model takes into consideration several factors for establishing the IQ of the information extracted from every source, introducing a percentage coefficient that we named IQA (information quality assessment), as

---

[18]We cannot forget the noise coming with the data flow, which must be carefully identified and removed. Of course, such a process has to be performed very cautiously since it may cause the definitive loss of precious data.

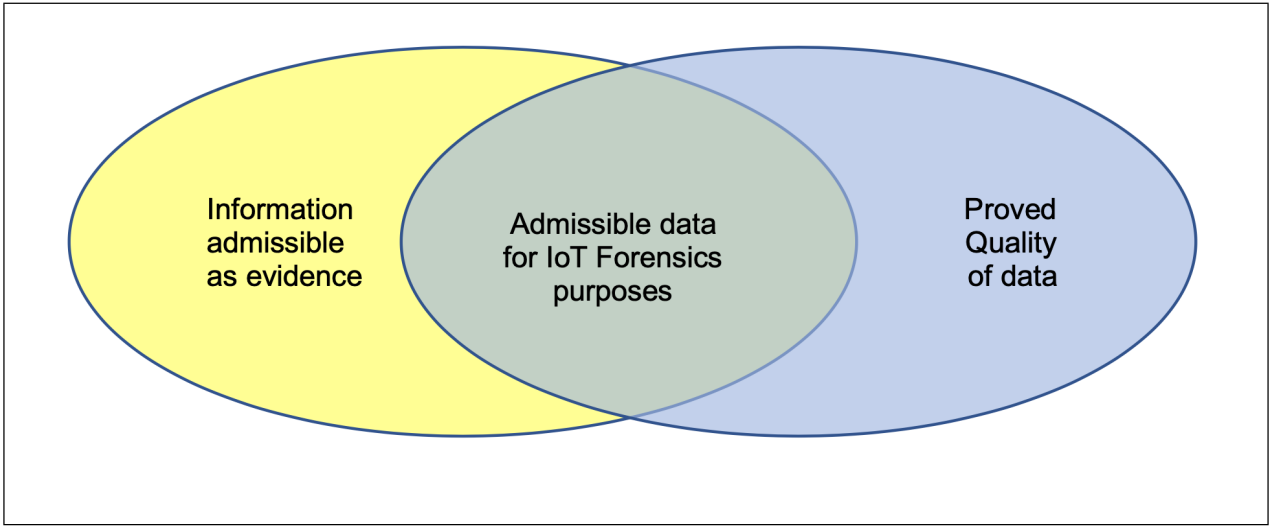

Figure 2: Requirements for IQ assessment. To be admissible for forensics purposes, the acquired information has to be, at the same time, readable and understandable as a digital data, and acquired and stored with respect to the best practices in the specific forensic field.

follows:
$$IQA = \frac{\sum_{i=1}^{n}(DTC_i + DST_i + CS_i + CM_i + SR_i + PC_i + TDA_i + OT_i + OS_i)}{9n} \times 100 \qquad (1)$$

where:

- $i$ = i-th device;
- DTC = device technical status;[19]
- DST = device security status (confidentiality, integrity, availability, ...);[20]
- CS = cloud service security status;[21]
- CM = cloud service manipulation of raw data;[22]
- SR = source reliability;[23]
- PC = privacy (GDPR) compliance;[24]

---

[19] The fact that the device is working properly (both hardware and software) and the currency of the software.

[20] The level of protection deployed in the device, both in hardware and software.

[21] The level of cyber protection of the cloud where data are stored, even temporarily, and of the transmission of data to and from the cloud.

[22] The possibility that in the cloud the data are processed and changed somehow.

[23] The possibility that the observer possesses additional information about the source which originates the data, i.e. the IQ level obtained in a previous investigation of data originating from the same source.

[24] The compliance with data protection regulations, for example in terms of data protection by default or by design. This requirement is related to the GDPR in the EU, yet of course it depends on the fact that in the legal system taken into consideration there is specific legislation.

- TDA = technical data accessibility;[25]
- OT = observer technological advancement;[26]
- OS = observer skills;[27]
- allowed values are all decimal between 0 = "bad" and 1 = "good";

Although such a model is more exploitable and complete than the one described in Figure 2, it is prone to both practical and theoretical objections:

1. It may not be possible to define each factor for all types of IoT devices. For example, the one named as CS seems to imply that the device is always connected to its environment, but some devices can also (or exclusively) be used offline;

2. The data needed to define these variables may not be accessible, not only technically, but because of the restrictions imposed by the companies that own them. For example, let's suppose that an investigation in Italy requires data related to a journey of a car produced by a foreign company. Let's think that, for some reason, the only copy of these data are stored in a server located the country where the company is based: it could be very complicated to reach such a source and collect data from it;

3. The terms of (1) belong to heterogeneous aspects, some technical, others mostly connected to the human factor, or depending on abstract concepts or that can be changed by the legislator. Therefore, it is arbitrary to estimate their mutual interactions, thus their contribution to (1). Questions like the following could arise: is it right that an increase of 0.1 in the device technical status has the same influence on (1) as the increase of 0.1 in the privacy (GDPR) compliance? What does it mean to increase, i.e. 0.15, in one of the items? Is the idea of "good" OS the same in Italy and the other countries?

4. In some cases it could be useful to set up an IN/OUT threshold value for IQA, a value under which any data can be discarded. This is an investigative choice that should be carefully considered and clearly explained.

All the aspects of the acquired data deserve to be explored in depth, without being mixed, and maybe confused or underestimated. In addition, thus giving an answer to objection number 4), we can't forget that in a forensic scenario all information must be taken under consideration, regardless of its quality, so dividing data depending on their belonging to one of the three categories assures us that we will not waste any details.

---

[25] The availability of technical specifications regarding the device and the format.

[26] The analytical tools used by the observer to collect and process evidence.

[27] The degree of experience, abilities, and skills achieved by the observer before the observation.

We suggest that the issue raised in point 3) can be tackled with the following precautions:

**a)** Pondering each element to add in the formula, depending on the contribution of the item to the whole evaluation;

**b)** Separating the items into different areas. Consequently, for every group of IoT evidence we would have different tags, yet scores can remain separated.

Concerning this last observation, we can divide the factors of the formula accordingly with two different kinds of taxonomies, whose adoption depends on the utility that it is more suitable in the trial. Of course, such a choice is not mutually exclusive, since in the trial every aspect can be used in argumentation. Indeed, we can highlight the kinds of information ("as", "about", or "for" reality) or the layers involved, as explained in the following paragraphs.

The two different kinds of taxonomies allow us to fine-tuning the model provided above into two opposite directions. In the first one we can aggregate the IQA factors into the four categories shown in Table 3, thus being able to split (1) in three formulas, each of which takes under consideration a different category of the "Philosophy of Information":

$$IQA_I = \frac{\sum_{i=1}^{n}(DTC_i + DST_i + CS_i)}{3n} \times 100 \qquad (2)$$

$$IQA_{II} = \frac{\sum_{i=1}^{n}(CM_i + SR_i)}{2n} \times 100 \qquad (3)$$

$$IQA_{III} = \frac{\sum_{i=1}^{n}(PC_i + TDA_i + OT_i + OS_i)}{4n} \times 100 \qquad (4)$$

where:

- $IQA_I$ = information *as* reality
- $IQA_{II}$ = information *about* reality
- $IQA_{III}$ = information *for* reality

In the latter point of view, (1) can be shaped considering that an IoT environment is characterized by the presence of different layers: physical perception layer, which includes sensing or moving; network layer, encompassing processing

| Categories | Philosophy of information | Requirements |
|:---:|:---:|:---:|
| Intrinsic | Information *as* reality (relevance) | DTC |
| Contextual |  | DST |
|  |  | CS |
| Representational | Information *about* reality (uncertainty) | CM |
|  |  | SR |
| Accessibility | Information *for* reality (accountability) | PC |
|  |  | TDA |
|  |  | OT |
|  |  | OS |

Table 3: Synopsis of IQ requirements and information categories.

and transmission; and application layer, which corresponds to the services provided [48].[28]

Each factor can be considered under three aspects, depending on the layer taken into consideration. As an example, DST – the technological security of the single device – can be evaluated under the physical aspect (i.e. protection case), the network security (i.e. encrypted transmission), and the application layer (i.e. password-protected interface).

Therefore, it is possible to draw the synopsis of Table 4[29], that drives us towards the definition of three other formulas that allow us to take under consideration this

---

[28]Such layers are defined as follows[76]: *"A physical perception layer that perceives physical environments and human social life, a network layer that transforms and processes perceived environment data and an application layer that offers context-aware intelligent services in a pervasive manner"*

[29]Of course, OT and OS are related to the observer, which is a human being. This means that the factors represent the technical apparatus deployed by the consultant and the skills acquired in each field. For example, a consultant could have a very advanced tool for device analysis (OTp), but not the same ability to guarantee the security of the connection (OTn) or the skills to evaluate how services are provided (OTa).

| Categories | Requirements / layers | | |
|---|---|---|---|
| | **Physical perception layer** | **Network layer** | **Application layer** |
| Intrinsic | DTCp | DTCn | DTCa |
| Contextual | DSTp | DSTn | DSTa |
| | CSp | CSn | CSa |
| Representational | CMp | CMn | CMa |
| | SRp | SRn | SRa |
| Accessibility | PCp | PCn | PCa |
| | TDAp | TDAn | TDAa |
| | OTp | OTn | OTa |
| | OSp | OSn | OSa |

Table 4: Synopsis of IQ and IoT layers.

different approach:

$$IQA_p = \sum_{i=1}^{h} \Big(\frac{\sum_{i=1}^{m_{p_i}} f_{p_j}}{m_{p_i}}\Big) \quad (5)$$

$$IQA_n = \sum_{i=1}^{h} \Big(\frac{\sum_{i=1}^{m_{n_i}} f_{n_j}}{m_{n_i}}\Big) \quad (6)$$

$$IQA_a = \sum_{i=1}^{h} \Big(\frac{\sum_{i=1}^{m_{a_i}} f_{a_j}}{m_{a_i}}\Big) \quad (7)$$

where:

- $h$ represents the number of devices taken into consideration in the investigation;
- $i$ = i-th device;
- $m_{p_i}$, $m_{n_i}$ and $m_{a_i}$ represent the number of factors that can be evaluated, respectively, in the physical, network, and application layer, for the i-th device. This is because, as clarified above, not all factors of Table 4 have to be present in the considered formula.
- $f_{p_j}$, $f_{n_j}$ and $f_{a_j}$ represent the single factor of Table 4 that can be evaluated (i.e. DTCp).
- $IQA_p$ = information of physical perception layer;

- $IQA_n$ = information quality of network layer;
- $IQA_a$ = information quality of application layer.

The considerations set out in this Section, make us confident of being able to make the "leap" towards operational realities, and to address the difficulties that surely could emerge in an investigation. In the next Section, we will expose two different case studies, that allow to appreciate how the terms abstractly defined in (1), and reused in (2), (3) and (4), could be practically evaluated and exploited.

# 6 Use cases of IQ in the IoT environment

As an application of all the above formulas in a real scenario, in the following section we propose two operational situation. Although both of them regard the issue of Information Quality Assessment, the first example is focused on the interpretation of the abstract terms in (1) in case of only one kind of digital evidence, still images and videos[30], whereas the latter, also proposed in [31], shows the powerful of the graphic visualization of the propose IQA assessment for an heterogeneous group of devices.

## 6.1 Case Study – 1: IQA assessment of images & video.

*On a crime scene, a video surveillance system were seized by the law enforcement agents. Subsequent investigations proved that the recorded footages wasn't saved on site, but instead uploaded in real time on a server located in a foreign country. The rogatory procedures for the acquisition of the data allowed law enforcement agents to retrieve the requested source of evidence. Investigators need to assess IQA of the obtained digital evidences applying (1) – (4).*

Before facing the exposed scenario, we move for a while from the meaning of IQ discussed in this paper, and we briefly try to focus upon the problem of determining the quality of a digital image, first by considering it "only" as a stand-alone digital representation of the real world, and then as a source of evidence that has to be acquired for a trial. To grasp the concept of digital image quality is not easy, due to the fact that it is not only a translation of the real world with "zeroes" and "ones" but also, or maybe mainly, a way to transmit emotions, a moment in life, and, of course, information. In the scientific literature this notion has various interpretations [8]:

- The subjective impression of **how well image content is rendered** or reproduced[63];

---

[30]One of the most useful sources of evidence in trials consists of visual documents [3]

- The integrated **set of perceptions** of the overall degree of excellence of an image[24];

- An impression of its merits or excellence as **perceived by an observer** neither associated with the act of photographing nor closely involved with the subject matter depicted[66];

- The perceptually weighted combination of all visually significant attributes of an image when considered in its marketplace or application ('Camera Phone Image Quality (CPIQ)-Phase 1-Fundamentals and Review of Considered Test Methods (v. 1.10)' 2007).

Other definitions are given from different projects and points of view[31]. In addition, we also wish to propose our first version of a definition of image quality, which is the following:

**Definition 1**: *The quality of a digital image is represented by its "closeness", both numerically and semantically, to the real scene it was intended to represent, considering one (or more) proper metric(s).*

The background of Definition 1 is that the digital image formation pipeline, described in detail in the literature [53, 32], possesses all the steps shown in Fig. 1, translated for the specific field as in Fig. 3. For this reason, we intend the digital image quality as being inversely proportional to the "difference" (defining a proper measure of distance in this interdisciplinary field is absolutely not a trivial task) between the real scene and its digital version. In other words, asking if an image has a "good" quality means asking "how well" the image allows to reconstruct all aspects of the shot scene, both objective and subjective. Although complete and exhaustive for an image not to be used as an evidence in a trial, the kind of quality requested in Definition 1 is not enough robust for a forensic scenario, as we are going to clarify in the following.

Image/Video Forensics is a part of Multimedia Forensics [9], which is a forensic science that aims to validate the authenticity of images and videos by recovering information about their history [60] for reasons connected to an investigation or a trial, where they appears as evidences. This implies that in the pipeline devoted to define the quality of an image or a video in a forensic context, we must also consider if this file has been handled with respect to the "chain of custody" described by the best practices in the field [40]. This means paying attention to:

- Where the image comes from;

---

[31]Among others, see [2] and [78]

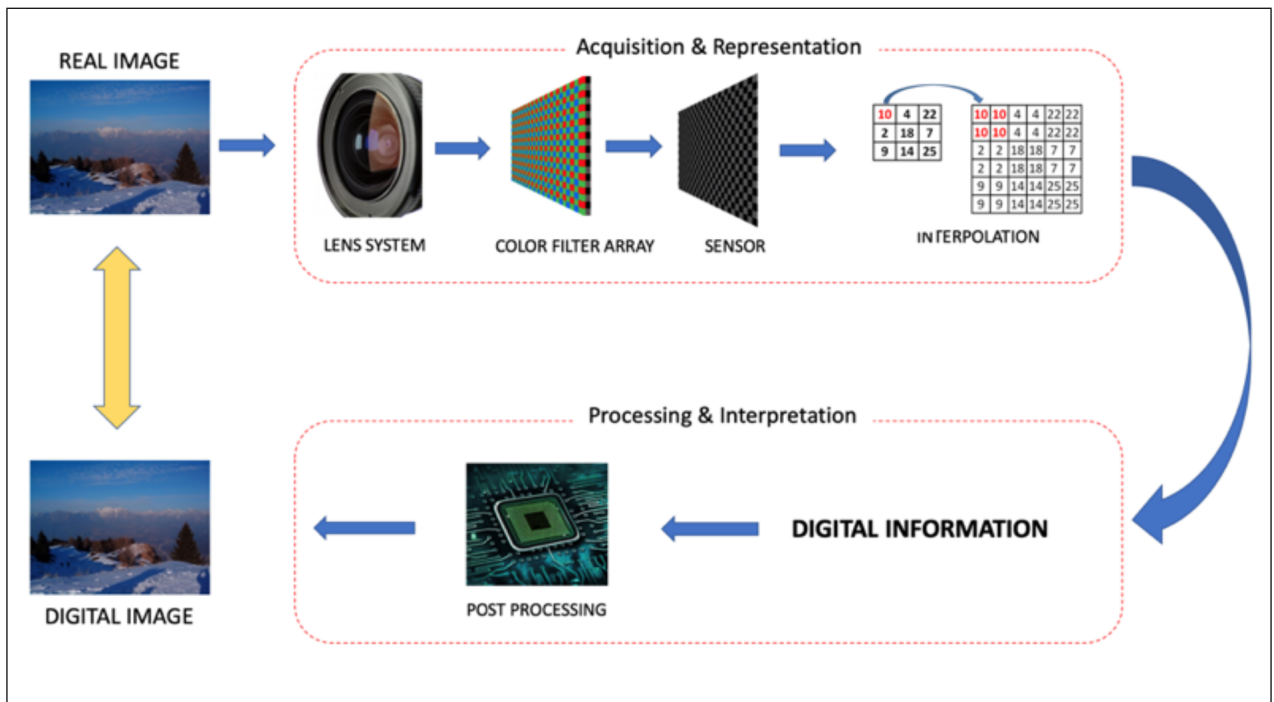

Figure 3: Steps of the formation of an image with reference to Figure 1. The real scene and its digital representation in an image are as close as the error (measured with a proper metric) between them is small.

- How the image has been acquired;

- How it has been manipulated during the analysis;

- How is has been preserved before and after the analysis.

We can also argue that definition (1) raises another issue, since it is not operational. Indeed, it is not suitable to denote the complexity of forensics procedures as described above, being unable to include a multidimensional representation of the device, its exchanges of information with its environment and the extraction of data made by the observer. A step forward is needed.

Therefore, depending upon this different point of view, the definition of digital image quality in a forensic context changes in:

**Definition 2**: *The quality of a digital image considered as evidence in a forensic process is represented by its "closeness", both numerically and semantically, to the real scene it was intended to represent, considering proper metrics, and the recommendation exposed in the best practices about the identification, collection, acquisition, and preservation of digital evidence.*

Assessing "Information Quality" in IoT ForensicsAfter these considerations, recalling the operational question raised above and with reference to all the terms of (1), answering to the query exposed in the begin of this section having in mind definition (2) means being able to complete the following checklist:

| Items | Meaning in the case of study |
|-------|------------------------------|
| DST | Is the hardware set up, wired, and maintained taking into account the security rules corresponding to the best practices? <br> Is the software devoted to managing the system provided with proper and up-to-date antivirus protection? <br> Are the accesses to the system properly logged? |
| CS | Is easily obtained the information about the channel towards data are broadcasted? <br> Is the entity from which data originate certified and reliable? |
| CM | Are data stored, even for a short period of time, in a repository that could be totally or partially accessible by some agent? |
| SR | Does the observer possess, or able to obtain, additional information about the source where the data originate? |
| PC | Are the footage and related metadata saved and preserved following the most recent GDPR precepts? |
| DTC | What is the technical status of the hardware part of the system? <br> Is the software part of the system updated to the latest release? |
| OT | Is the observer recently acknowledged in some way as a valuable operator in the field? |
| TDA | Are the technical specifications regarding the device and the format of the obtained data easily available, or provided directly by the owner of the data? |
| OS | What is the expertise of the observer as a digital Forensics investigator? |

Table 5: Checklist of issues emerging from IoT Forensics.

Facing an operative scenario with the exposed mathematical model, being able to assess numerical values to every term of (1) is crucial. Every type of evidence requires different approaches. In this table the checklist needed to calculate the IQAs from (1)-(4).

It is remarkable that definition (2) of image quality embraces a more pragmatic perspective. In adopting it, forensics experts need to be aware that this needs to fulfill a methodological further pledge of transparency – which can be reconnected with an obligation of "accountability", as stated above – thus requiring the development of an argumentative strategy aimed at avoiding misinterpretations.

## 6.2 Case Study – 2: IQA assessment for a set of digital evidences.

*On a crime scene, a set of IoT digital devices were seized. Specifically, a smartphone; 2) the SIMCard inside of it; 3) a drone; 4) a smartwatch; 5) a laptop pc; 6) a smartTV. A Digital Forensics expert, before analyzing in depth these sources of evidence, must ascertain the IQ of every single device both individually and globally.*

Comparing IQs of the devices and evaluating the overall IQ taking under consideration all the terms that compose (1) could be a difficult task. Indeed, the device manufacturers may not (or at least not yet) have made public the requested technical information or, even if released, these data could be not as detailed as necessary. In addition to that, information inside each device could be organized and stored in different ways, depending on the policies of the respective manufacturer. The level of these evaluations should be similar to what exposed in [16], [13] ad [57], where the file system, the shape and the format of the log files and other useful forensic clues are exposed in case of a drone, a smartTV and a smartwatch. After this kind of deep analysis, we could fill a table as Table 7 below[32], and then implementing (1), (2), (3) and (4), and thus generating a set of charts that allows to better insight the IQ of the examined sources of evidences. For the test we considered the following devices:

1. Smartphone Huawei model ALE-L21 (P8 Light), with Android 6.0, 2 Gb RAM, CPU Octa-core 1.2 GHz, kernek version 3.10.86-g33ff982;

2. Nano SIMCard 4G Telecom Italia year 2017;

3. The same model proposed in in [54];

4. The same model proposed in in [13];

5. IBM Thinkpad Edge E30, o.s. Windows 10, 8 Gb RAM, Intel i5 processor;

6. The same model proposed in in [57].

By applying (1), (2), (3) and (4) to all devices, with the data exposed in Tab.6 as input, we obtain the results listed in Tab.7: the IQA of the set of all seizured devices is about 62%, the device nr.4 is the one achieving the best result in terms of Information Quality, whereas device nr.5 bears the worst performance:

---

[32] The exact numerical value inserted in this example for every entry of Table 7, is not so crucial for the purpose of this paper, since were calculated after an evaluation made by the authors, and they could be subjected to significant variation depending upon a lot of reasons, e.g. changes in the laws, technical improvements, enhanced technical ability, and so on. Their purpose is allowing the calculation of the requested IQs and the graphical representation of the results.

| device 1 smartphone | | device 2 SIMCard | | device 3 drone | | device 4 smartTV | | device 5 pc laptop | | device 6 smartwatch | |
|---|---|---|---|---|---|---|---|---|---|---|---|
| DTC | 0,56 | DTC | 0,93 | DTC | 0,97 | DTC | 0,91 | DTC | 0,39 | DTC | 0,89 |
| DST | 0,62 | DST | 0,12 | DST | 0,48 | DST | 0,16 | DST | 0,30 | DST | 0,82 |
| CM | 0,34 | CM | 0,17 | CM | 0,76 | CM | 0,60 | CM | 0,58 | CM | 0,85 |
| SR | 0,48 | SR | 0,76 | SR | 0,50 | SR | 0,98 | SR | 0,56 | SR | 0,18 |
| PC | 0,77 | PC | 0,82 | PC | 0,77 | PC | 0,19 | PC | 0,00 | PC | 0,98 |
| TDA | 0,55 | TDA | 0,60 | TDA | 0,21 | TDA | 0,44 | TDA | 0,26 | TDA | 0,65 |
| OT | 0,84 | OT | 0,07 | OT | 0,89 | OT | 0,80 | OT | 0,07 | OT | 0,89 |
| OS | 0,26 | OS | 0,88 | OS | 0,45 | OS | 0,64 | OS | 0,79 | OS | 0,31 |

Table 6: Numerical values of terms in (1), as evaluated by the authors for devices 1 – 6.

| | | | | |
|---|---|---|---|---|
| $IQA_I$ = 61,96 % | $IQA_{III}$ = 54,74 % | $IQA_{device2}$ = 59,49 % | $IQA_{device4}$ = 89,79 % | $IQA_{device6}$ = 66,68 % |
| $IQA_{II}$ = 56,30 % | $IQA_{device1}$ = 54,37 % | $IQA_{device3}$ = 63,73 % | $IQA_{device5}$ = 38,19 % | $IQA_{tot}$ = 62,04 % |

Table 7: Results of the application of (1), (2), (3) and (4) to the seizured devices, using the data exposed in Tab.6.

The IQ can be clearly visualized also by a set of *radar* chart, which offers an immediate representation. In Figure 4, a set of evaluations is showed, considering both the total of the acquired staff and the single device. Subfigure a) represents a model of the best result that can be achieved: all the elements that compose the evaluation are at the maximum level, so the polygon is completely surrounded by the blue line. Subfigure b) shows at the same time the IQA of all the examined devices, and allows appreciating immediately the best result of devices nr.4 already highlighted. Subfigure c) shows together the IQA calculate with (2), (3) and (4), whereas in subfigures from d) to i) the performances of every single device is represented. Also from the comparison between these latter group of images, the peaking values of device nr.4 clearly appears among the others.

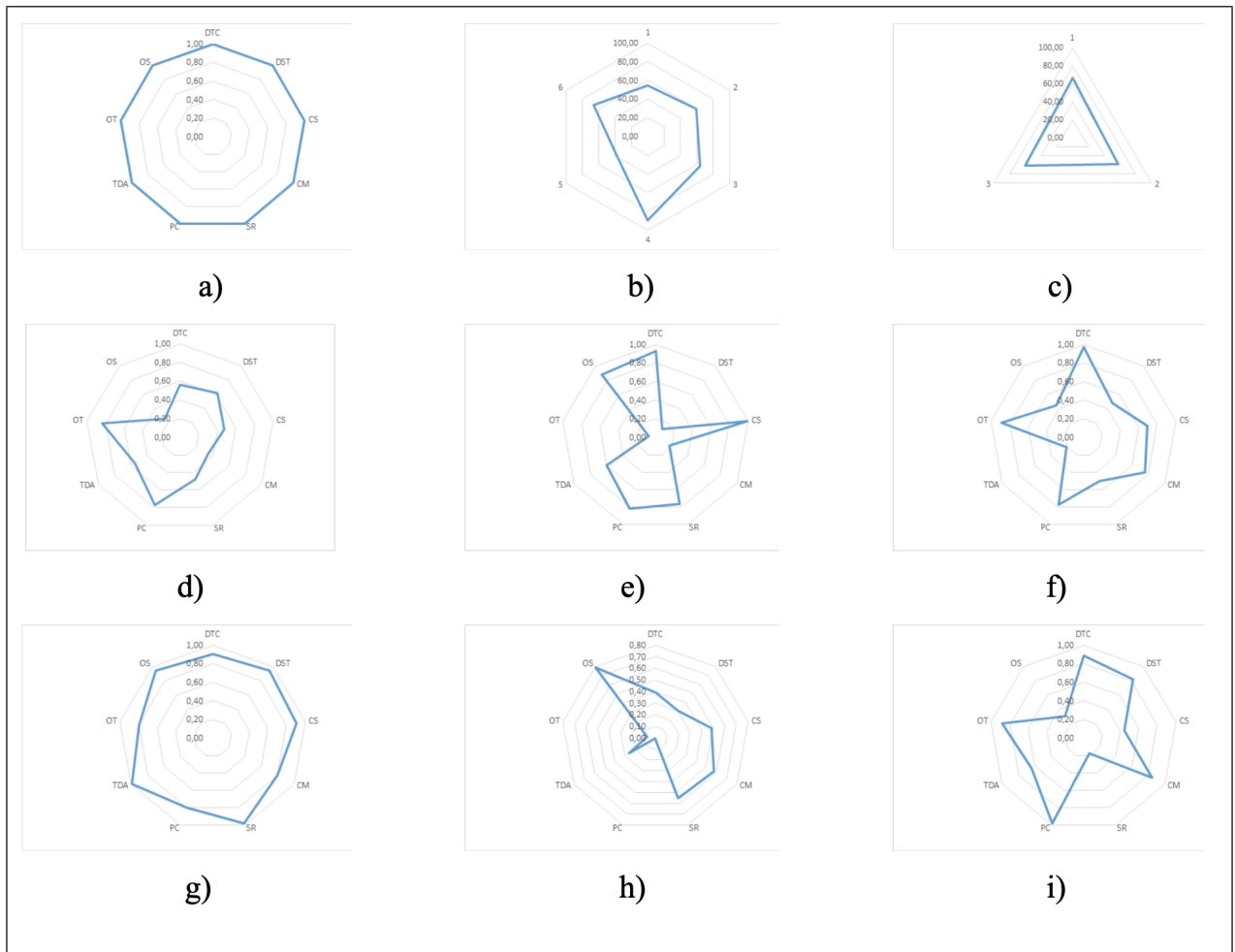

Figure 4: Graphic visualization of the outcomes of Case Study 2. The results of the IQA calculated by (1), (2), (3) and (4) become more intelligible with the help of this kind of charts, where the bigger the part of the inner figure is surrounded, the best is the achievement. In the above representation, a) was given as a model, and represents an example of the best result that a certain evaluation could achieve, since all the evaluated terms are at the higher level; b) is the IQA of all the devices showed together, that allows to highlight how, in the considered case work, the device 4 is the one with the best performance; c) shows at the same time the $IQA_I, IQA_{II}$ and $IQA_{III}$; d) - i) are the charts referred to every term of the formulas of every devices, respectively 1 - 6. Also in this comparison confirms that devices nr.4 obtains the best result.

# 7 Conclusions

Civil and criminal proceedings can be side-tracked from reaching their fair conclusion by mistakes or misinterpretations made by several actors: judges, defendants, prosecutors, police officers, consultants, witnesses. In this, digital evidence has become more and more important not only because it is increasingly widespread but also because of the ever-increasing skills required to manage it. It is crucial to improve the efficiency and the efficacy of evidentiary processes and to raise awareness of the peculiar nature of digital evidence for all the professionals involved.

Of the many challenges currently faced by digital Forensics, IoT technologies are of most concern, not only for practical but also for theoretical reasons.

From a theoretical perspective, many questions need to be deepened. Aside from the discussion of DQ and IQ approaches, it is remarkable that IQ is shaped differently depending on the level of complexity addressed (human to human, human to machine, machine to machine). Additionally, the problem of quality should be considered a fundamental issue in the design of IoT devices and ecosystems. Ethical concerns in this field should focus on the many facets of quality.

From a practical perspective, one of the main issues is to comply with the requirement of relevancy of digital evidence. Indeed, in Digital Forensics, *a priori* discarding of data is not recommended, because of the risk of deleting important details. Yet Forensics experts need to understand which type of data must be acquired and how it should be manipulated. From this view, IQ could be considered a kind of metadata: information *about* information. In a nutshell, to analyse IoT evidence, it is essential to track data from the original source to the data repository. IoT Forensics simply follows the path of information going backwards.

IoT Forensics is one of the new subareas of digital Forensics [77, 37], and very few studies have been carried out with the specific aim of modelling theoretical approaches [55, 15, 5].

In this contribution, we drew a theoretical framework and introduced a model for assessing IQ in IoT Forensics. Our approach, which we expressed in a general formula and three applications – one for each layer – aims to describe the complexity of IQ in the IoT environment. To the best of our knowledge, this is the first attempt to provide a mathematical framework for a purely theoretical issue, that of the legal argumentation for evaluating evidence in trial, in the case of the IoT environment[33]. Further steps more steps need to be made along this path[34].

---

[33] This involves technical evaluation of the quality of a new kind of evidence, which could be acquired from different and heterogeneous sources, maybe without a law enforcement agent being physically on site

[34] After a previous introduction [19], we made a further step [31] but many more are required

First, we aim to implement our approach, exploring the possibility of creating a database for every data type, thus allowing us to obtain robust statistics, and in so doing, developing an approach which involves Gaussian mixture models theory [62, 65, 70].

Second, in the future we aim to implement our model integrating the ISO/OSI level of complexity by adding four more layers, based on the interaction with the human factor, which is pivotal in Forensics, since the evidence has to be discussed in courts.

Third, we aim to provide further details on the features of the observer, now generally expressed as OS.

Fourth, it is important to assess the impact of this approach on a legal aspect, evaluating if the rights and fundamental freedoms of individuals – indicted and victims – could be endangered by its implementation. It is obvious that most threats could emerge in the processing of personal data, thus posing risks for discrimination and unfair treatment.

Fifth, we believe that benchmarks should be established in order to assess the reliability of our model. To create such benchmarks, the possibility of including data from previous evaluations could be explored. For example, one could reuse data concerning the analysis of certain models of devices (i.e. the Samsung "smart tv" model UE65NU7400U) in order to fine-tune certain variables (i.e. accessibility) and, in a more sophisticated model, integrate such assessment with technical specifications provided directly by manufacturers and feedback from the community of forensic experts. Perhaps in a relatively near future devices could be classified not only in terms of energy consumption but also by their cybersecurity or forensic trustworthiness.

Assessing "Information Quality" in IoT Forensics